\documentclass[aps,prl,twocolumn,groupedaddress,nofootinbib,showpacs]{revtex4}
\usepackage{graphicx} 
\usepackage{amsmath}

\begin{document}

\title{Simulating electrochemical systems by combining the finite field method with a constant potential electrode}

\author{Thomas Dufils$^1$, Guillaume Jeanmairet$^1$, Benjamin Rotenberg$^1$, Michiel Sprik$^2$ and Mathieu Salanne$^1$}
\affiliation{$^1$ \small Sorbonne Universit\'{e}, CNRS, Physico-chimie des \'Electrolytes et Nanosyst\`emes Interfaciaux, PHENIX, F-75005 Paris, France}
\affiliation{$^2$ \small Department of Chemistry, University of Cambridge, Cambridge CB2 1EW, United Kingdom}

\date{\today}

\begin{abstract}
A better understanding of interfacial mechanisms is needed to improve the performances of electrochemical devices. Yet, simulating an electrode surface at fixed electrolyte composition remains a challenge. Here we apply a finite electric field to a single electrode held at constant potential and in contact with an aqueous ionic solution, using classical molecular dynamics. The polarization yields two electrochemical interfaces on opposite sides of the same metal slab. While the net charge on one electrode surface is the opposite of the net charge on the other, maintaining overall charge neutrality of the metal. The electrode surface charges fluctuations are compensated by the adsorption of ions from the electrolyte, forming a pair of electric double layers with aligned dipoles.  This opens the way towards the efficient simulation of electrochemical interfaces using any flavor of molecular dynamics, from classical to first principles-based methods.
\end{abstract}

\maketitle

	Despite many advances over the past decades~\cite{callevallejo2012a,magnussen2019a}, the efficient simulation of full electrochemical cells at the molecular scale, using electronic structure based calculations, remains a daunting task. This is due to their slab structure, since the minimal experimental setup consists of an electrolyte between two electrodes. The system is generally simplified by simulating one interface only, but the main conceptual difficulty is to find a way to charge the electrode surface at fixed composition of the electrolyte. Several methods have recently emerged~\cite{bonnet2012a,melander2019a,nishihara2017a}, where the system is allowed to exchange electrons with a reservoir at fixed voltage (grand-canonical approach), but they all rely on the use of continuum descriptions for the electrolyte. These models, which are generally based on a Poisson-Boltzmann theory~\cite{nattino2019a}, remain mostly qualitative and an atomistic description would be preferable (this is also true because the solvent may actively participate to electrochemical reactions~\cite{le2017a,bouzid2018a}). This is almost impossible to do since it would be necessary to remove/insert ions to counterbalance the electrode charge fluctuations.

Here we propose an alternative route to simulate electrochemical cells. Our approach is based on the coupling of a finite field with a system consisting in an electrolyte and a single electrode. Finite fields methods, developed in the framework of the modern theory of polarization, consist in imposing a macroscopic field (electric field~\cite{souza2002a}, polarization~\cite{dieguez2006a} or electric displacement \cite{stengel2009a}) via an extended Hamiltonian accounting for the interaction between the system and the fixed field. They have recently been adapted and applied to the study of electrical double layers at solid/liquid interfaces, and more precisely charged \cite{zhang2016f} or polar \cite{sayer2017a} insulators/electrolyte interfaces. Electrochemical systems are by nature more complex since they involve metallic electrodes whose charge distribution is not fixed but depends on the surrounding medium and the applied potential. Due to the long simulation times related to the relaxation of the electrical double layer (which is typically longer than the nanosecond), we establish here a proof of concept by using a classical molecular dynamics (MD) setup. Indeed, even if metals can only be accurately described in the framework of quantum mechanics, models have been developed to reproduce the electrostatics in classical or mixed quantum/classical simulations~\cite{siepmann1995a,golze2013a}. In this contribution we extend the finite electric field method to an electrolyte interacting with such a model metallic electrode.

In the following, we focus on the description of the electrostatics, the Van der Waals interactions being represented by the conventional Lennard-Jones model. The charge density in any point of space is given by
	\begin{equation}
		\rho(\mathbf{r})=\sum_{i=1}^N q_i\delta({\bf r}-{\bf r}_i)+\sum_{j=1}^M q_{j}\eta^{3}\pi^{3/2}\exp\left[-\eta^{2}(\mathbf{r}-\mathbf{r}_{j})^{2}\right]
	\end{equation}	
	where the first term is the contribution of the  electrolyte, represented by a distribution of point charges with $q_i$  the partial charge of the atom $i\in[1,N]$ and ${\bf r}_i$ its position; $\delta$ is the Dirac distribution. The second term represents the atoms of the metallic electrode, in which each site $j\in[1,M]$ is immobile (with position ${\bf r}_j$) and carries a charge $q_{j}$ which is spatially distributed following a gaussian charge distribution  of width $\eta^{-1}$. In order to represent the metallic character of the electrodes, the latter charges are allowed to fluctuate in response to the electrolyte fluctuations and thus are part of the microscopic degrees of freedom~\cite{siepmann1995a,limmer2013a}. The Hamiltonian of the system is written as
\begin{equation}
H^{PBC}=K(\mathbf{p}^{N})+U(\mathbf{r}^{N},\mathbf{q}^M),\label{Hamiltonian_PBC}
\end{equation}

\noindent where $K(\mathbf{p}^{N})$ is the kinetic energy which depends on the ion momenta $\mathbf{p}^{N}=\lbrace \mathbf{p}_{1}...\mathbf{p}_{N}\rbrace$ and $U(\mathbf{r}^{N},\mathbf{q}^M)$ the potential energy, which depends on the ion positions $\mathbf{r}^{N}=\lbrace \mathbf{r}_{1}...\mathbf{r}_{N}\rbrace$ and charges of the electrode $\mathbf{q}^M=\lbrace q_{1}...q_{M}\rbrace$. We have appended a superscript PBC (periodic boundary conditions) to indicate that the electrostatic energies and forces are computed using standard Ewald summation to account. 
	All the atoms within an electrode are held at constant potential by enforcing the following condition on each atom
\begin{equation}
	\Psi^{PBC}_{j}=\dfrac{\partial U^{PBC}_C}{\partial q_{j}}=\Psi_{\rm  J}
	\label{eq:potential}
\end{equation}

\noindent where $\Psi_{\rm J}$ is the prescribed potential of electrode J to which the atom $j$ belongs and $U^{PBC}_C$ is the Coulombic contribution to the energy, given by:
\begin{equation}
	U^{PBC}_{C}=\dfrac{1}{2}\iint \dfrac{\rho(\mathbf{r}) \rho(\mathbf{r'})}{4\pi\epsilon_{0}\vert \mathbf{r}-\mathbf{r}' \vert} {\rm d}\mathbf{r}{\rm d}\mathbf{r}'.\label{Uc}
\end{equation}
	
	Formally, solving the set of self-consistent equations given by Equation \ref{eq:potential} is equivalent to minimizing $U^{PBC}_C-\sum_j \Psi_{\rm J}q_j$ with respect to the charges. Since this function is quadratic in the fluctuating charges $q_{j}$ (see Eq \ref{Uc}), the minimization can be efficiently performed with conjugate gradients. Note that we add an additional constraint by forcing the sum of the electrode charges to be null~\cite{gingrich2010a}.


\begin{figure}[!h] 
\includegraphics[width=\columnwidth]{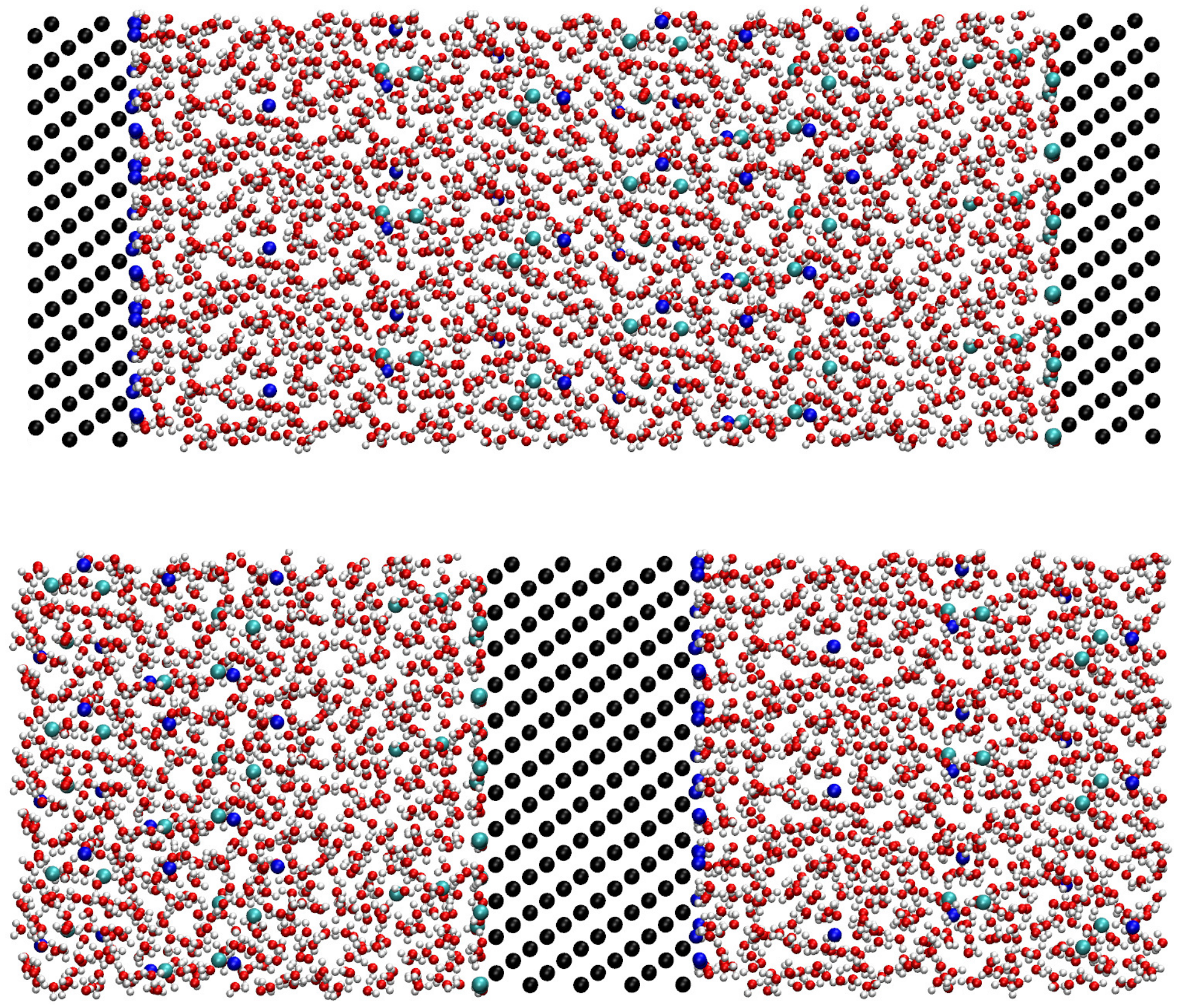}
	\caption{Top: electrolyte-centered supercell (2D PBCs), with fixed applied potential between two electrodes ($\Delta \Psi=\Psi_{\rm right}-\Psi_{\rm left}$). Bottom: conductor-centered supercell (3D PBCs), with a finite field ${\bf E}$ and a single electrode in which the potential of the atoms is set to the same value $\Psi_{\rm electrode}$.\label{ECSvsCCS}}
\end{figure}

	The minimal components of an electrochemical cell  are two electrodes and an electrolyte between them. In classical MD, the computational cost is not prohibitive so it is relatively easy to simulate complete systems instead of a single electrode surface. The conventional setup to simulate such systems is illustrated in the top panel of Figure \ref{ECSvsCCS}. In the following we will refer to this system as the electrolyte-centered supercell (ECS). It is simulated using the 2D Ewald summation~\cite{kawata2001a, reed2007a} since the two electrodes are held at different potentials ($\Psi_{\rm J}=\Psi_{\rm left}$ or $\Psi_{\rm right}$). We note $x$ and $y$ the two directions along which PBCs are used in this setup.

	Finite field ($\mathbf{E}$) simulations can be performed using the extended Hamiltonian introduced by Stengel and Vanderbilt in \cite{stengel2009a} and is written as:
\begin{equation}
H_{E}=H^{PBC}-\Omega ~\mathbf{P}\cdot\mathbf{E},\label{Hamiltonian}
\end{equation}
\noindent where $H^{PBC}$ is the Hamiltonian defined by Eq \ref{Hamiltonian_PBC}, $\Omega$ is the volume of the supercell and $\mathbf{P}$ the polarization per unit volume. In the modern theory of polarization, the dipole moment of a unit cell involving infinite periodic systems is viewed as a multivalued quantity, since it depends on the choice of the position of the periodic boundaries. Nevertheless, this is not a significant issue since only differences in polarization matter in the dynamics and in the calculation of physical properties, in practice via the itinerant polarization \cite{caillol1994a}:
\begin{equation}
	\mathbf{P}_{\rm itinerant}(t)=\mathbf{P}_{\rm itinerant}(0)+\dfrac{1}{\Omega}\sum_{i=1}^N q_{i}\Delta \mathbf{r}_{i}(t),
\end{equation}

\noindent where $\Delta\mathbf{r}_{i}$ is the displacement of the atom between time $t=0$ and $t$ for the "unfolded" trajectory, i.e. not taking jumps in position (hence polarization) across the periodic boundaries. The system must be periodic in the direction in which the finite electric field is applied,  which implies  the use of 3D PBCs. A field $E$ corresponds to a drop of Poisson potential across the cell $\Delta\Psi=-E L_{z}$ where $L_{z}$ is the length of the box in the direction of the field. From the practical point of view, a consequence of the use of 3D PBCs is that now the two electrodes of the ECS necessarily merge, yielding a single electrode at fixed potential $\Psi_{\rm J}=\Psi_{\rm electrode}$. The simulation cell can then be represented with the electrode at its center, yielding the conductor-centered supercell (CCS) shown in the bottom panel of Figure \ref{ECSvsCCS}. Contrarily to the constant applied potential simulations, it is now the presence of the finite field that induces the polarization of the electrode and potential drop at the two electrode/electrolyte interfaces. 

\begin{figure*} 
\includegraphics[width=\textwidth]{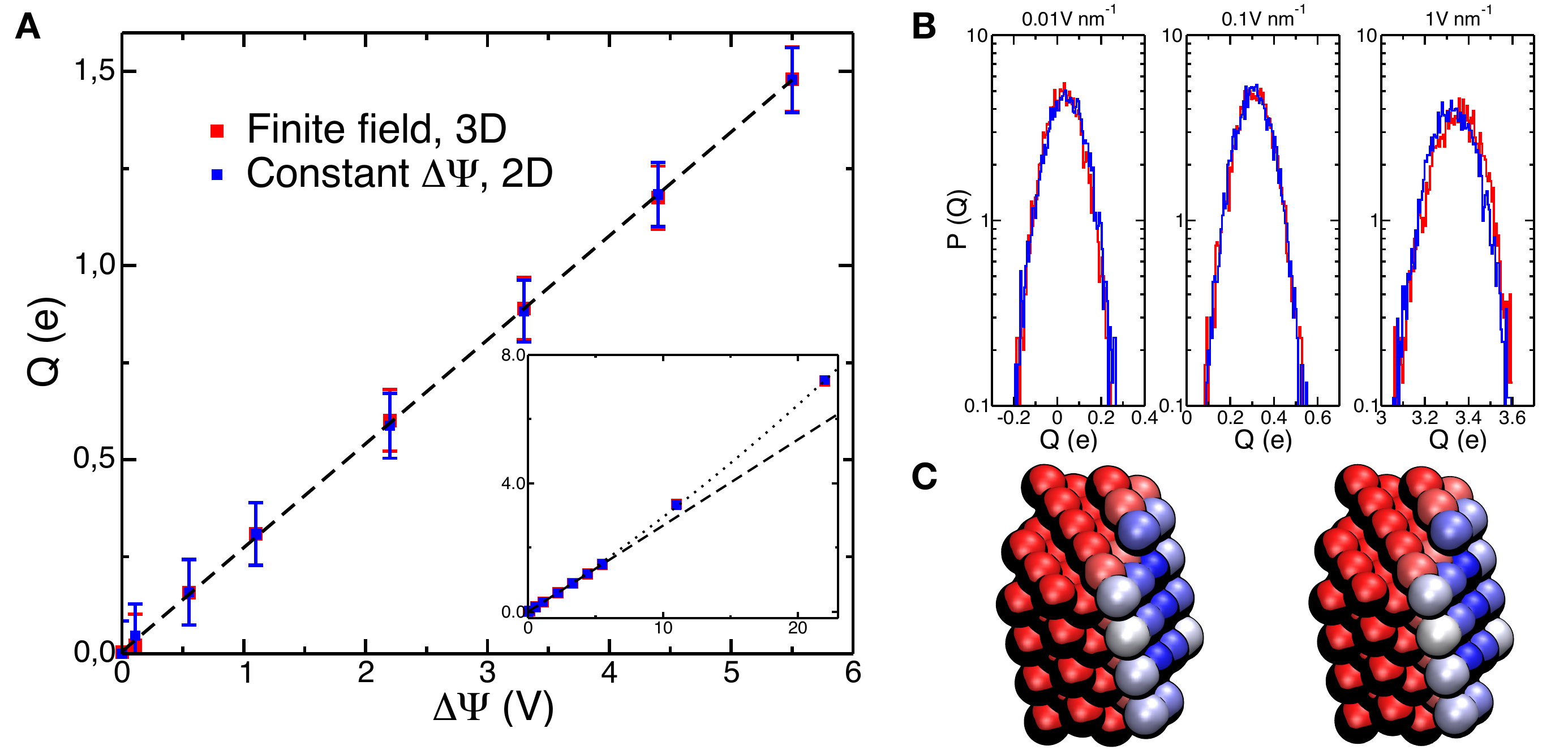}
	\caption{The single electrode behaves as two distinct constant potential electrodes. A) Average of the total accumulated charge on the positive side as a function of the applied potential. The error bars correspond to the standard deviation of the charge distribution. B) Normalized distributions of the instantaneous total charge on the positive half-electrode for various finite electric fields ($E~=~0.01$, 0.1 and 1~V~nm$^{-1}$ from left to right, red lines) or the corresponding applied voltages ($\Delta \Psi~=~0.11$, 1.1 and 11~V from left to right, blue lines). C) Snapshot of the positive electrode for a finite field of 0.5~V~nm$^{-1}$ (left) and the corresponding applied potential of 5.5~V (right), for the same electrolyte configuration. The atoms are colored according to their instantaneous charge, between 0~e (red) and 0.1~e (blue).\label{validation-charge}}
\end{figure*}

Coupling fluctuating charges to model conductors with finite field simulations requires two important adaptations of both methods. On the one hand, the cell polarization includes a contribution from the fluctuating charges as:
\begin{equation}
	\mathbf{P}_{\rm conductor}(t)=\dfrac{1}{\Omega}\sum_{j=1}^M q_{j}(t)\mathbf{r}_{j}.\label{polar_electrode}
\end{equation}

This additional term does not depend on the position of the electrode inside the supercell since we enforce $\sum_j q_j=0$. On the other hand, the determination of the partial charges via the constraint of fixed potential includes an additional contribution to the electrostatic energy and potential due to the finite field. From the extended Hamiltonian (Eq \ref{Hamiltonian}) and the expression of the electrode contribution to polarization (Eq \ref{polar_electrode}), one obtains the generalization of Eq \ref{eq:potential} as:
\begin{equation}
	\Psi_{j}=\Psi^{PBC}_{j}-\mathbf{r}_{j} \cdot \mathbf{E}=\Psi_{\rm  J}
	\label{eq:potential_field}
\end{equation}

As an extension of the constant applied potential case, the self-consistent expressions given by Eq \ref{eq:potential_field} are now solved by minimizing $U_C^{PBC} -\Omega ~\mathbf{P} \cdot \mathbf{E} -\sum_j \Psi_{\rm J}q_j$ with respect to the charges. Since ${\bf P}$ is a linear function of the charges, the performances of the conjugate gradient minimizer are not affected by this setup.

In order to test this new approach, we simulate two systems, one at constant applied potential between two distinct electrodes and the other with the finite field method and a single fixed-potential electrode as shown on Figure \ref{ECSvsCCS}. The electrode(s) consist in a model structure made of a cubic crystal with the NaCl lattice constant,  with the (111) plane facing the liquid. The intermolecular interactions consist in electrostatic interactions and Lennard-Jones potentials using Lorentz-Berthelot mixing. The electrode sites Lennard-Jones parameters are the ones of Cl$^-$~ \cite{joung2008a}. The electrolyte consists in an aqueous solution of NaCl composed of 603 SPC/E water molecules~\cite{berendsen1987a} and 20 ion pairs~\cite{joung2008a}. The cross-sectional area is 2.20~nm$^{2}$ and the length along the $z$ axis $L_{z}=11$~nm. 

The simulations were performed with a timestep of 2~fs in the NVT ensemble at 298~K using a Nos\'e-Hoover thermostat with a coupling constant of 0.4~ps. The systems were equilibrated during 10~ns before a production run of 10~ns.

For the finite field simulations, we use the CCS configuration with 3D PBCs. The single electrode, which is made of 12 planes of atoms, is set at null potential and the field $E$ ranges from 0 to 2~V~nm$^{-1}$. For the constant applied potential simulations, we setup the ECS configuration with 2D PBCs. In this setup the two electrodes are of equal dimensions ({\rm i.e.} 6 planes of atoms each), kept under constant potentials $\Psi_{\rm left}$ and $\Psi_{\rm right}$ such that $\Delta \Psi=\Psi_{\rm left}-\Psi_{\rm right}=-E L_{z}$, using the same values for $E$ as above.  In both series of simulations, the value of the $\eta$ parameter for the gaussian charges has been set to 0.5052~\AA\ $^{-1}$ following ref. \cite{reed2007a}. Electroneutrality is enforced during the charge calculation process~\cite{gingrich2010a}.

A first validation is provided by comparing the polarization of the electrodes with the two setups. To do this we split the single electrode in the CCS setup in two parts. This is easily made since one side accumulates positive charge and the other is exactly opposite, while the centre is almost neutral. Firstly, the average accumulated charge $Q$ on the positive side is compared with the ECS setup on Figure \ref{validation-charge}A for a wide range of applied fields (potentials). The agreement between the two methods is excellent since the points are almost superimposed. This is true not only for voltages up to 6~V where the charge increases linearly with the applied potential (which reflects a constant differential capacitance~\cite{limmer2013a}), but also up to 20~V for which the variation is not linear anymore (see the inset). The probability distributions of the instantaneous values of this quantity (Figure \ref{validation-charge}B) are also identical, which shows that the two methods sample the same configuration space.  Finally,  Figure \ref{validation-charge}C illustrates the instantaneous electrode charges for a given electrolyte configuration. The two systems are indistinguishable, showing that even at the local scale the finite field method yields a correct representation of the electrode/electrolyte interface.

\begin{figure} 
\includegraphics[width=\columnwidth]{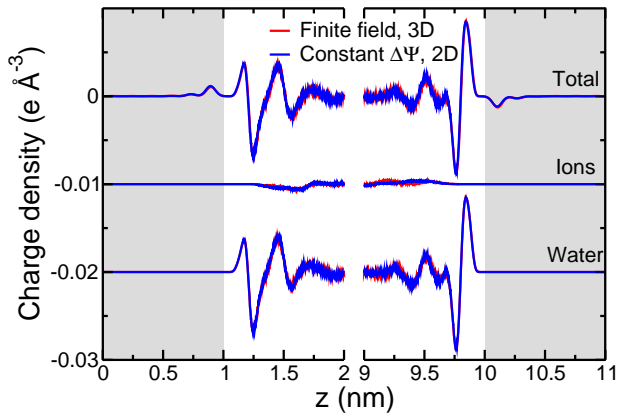}
	\caption{Charge density within the electrode and the electrolyte for a finite field of 0.1~V~nm$^{-1}$ or the corresponding applied potential of 1.1~V. The contributions of water molecules and ions to the total charge distribution are also shown; they were shifted by -0.01~e~\AA$^{-3}$ for a better readability. The grey-shaded areas correspond to the electrodes (the CCS was translated in order to match with the ECS setup).\label{validation-electrolyte}}
\end{figure}

When a finite field is applied to a bulk liquid, due to the PBCs the charges cannot accumulate in a specific region so that there is a net electric field in each point of the supercell. Here the presence of a blocking surface (the electrode) results in the formation of polarized layers on the electrolyte side. The polarization arises from two mechanisms: i) reorientation of the water molecules ii) local charge imbalance by accumulation of one ionic species and depletion of the other. The structure adopted by the liquid may be compared with the case of constant applied potential simulations. As shown on Figure \ref{validation-electrolyte}, the agreement is again very good for  the variation of the charge density and its splitting between water and ionic contributions across the simulation cell. This validates further the use of the single electrode in order to study electrochemical interfaces using 3D PBCs. Additional tests on the electric field and Poisson potential for several applied voltages are provided in the Supplementary Information; they all show the same level of accuracy.

In conclusion, we have demonstrated the possibility to simulate metal slabs with fluctuating surface charge and in the presence of an explicit electrolyte that counterbalances the charge. This is done through the combination of two methods: An applied finite field which polarizes the cell and a constant potential electrode that screens the field in bulk, leading to the formation of two independent electrochemical interfaces. The net charge on one electrode surface is the opposite of the net charge on the other, which maintains the overall charge neutrality of the metal slab. The electrode surface charge fluctuations are compensated by the adsorption of ions from the electrolyte. The method is validated through extensive comparisons with simulations using constant applied potential between two separated electrodes (and 2D PBCs). From the practical point of view, it is much easier to implement in classical MD packages since it avoids the introduction of 2D PBCs. From the computational point of view, the method is also more efficient: The simulation time is reduced by approximately 15~\%. Last, but not least, it could much be more easily applied to the case of {\it ab initio} molecular dynamics, which would open the door for the first-principles simulation of electrochemical reactions occurring at electrodes, in the presence of an explicit electrolyte.

\section*{Supplementary Information}

\subsection*{Constant potential and applied electric field}

\begin{figure}[!h] \centering
\begin{center}
\includegraphics[width=0.4\textwidth]{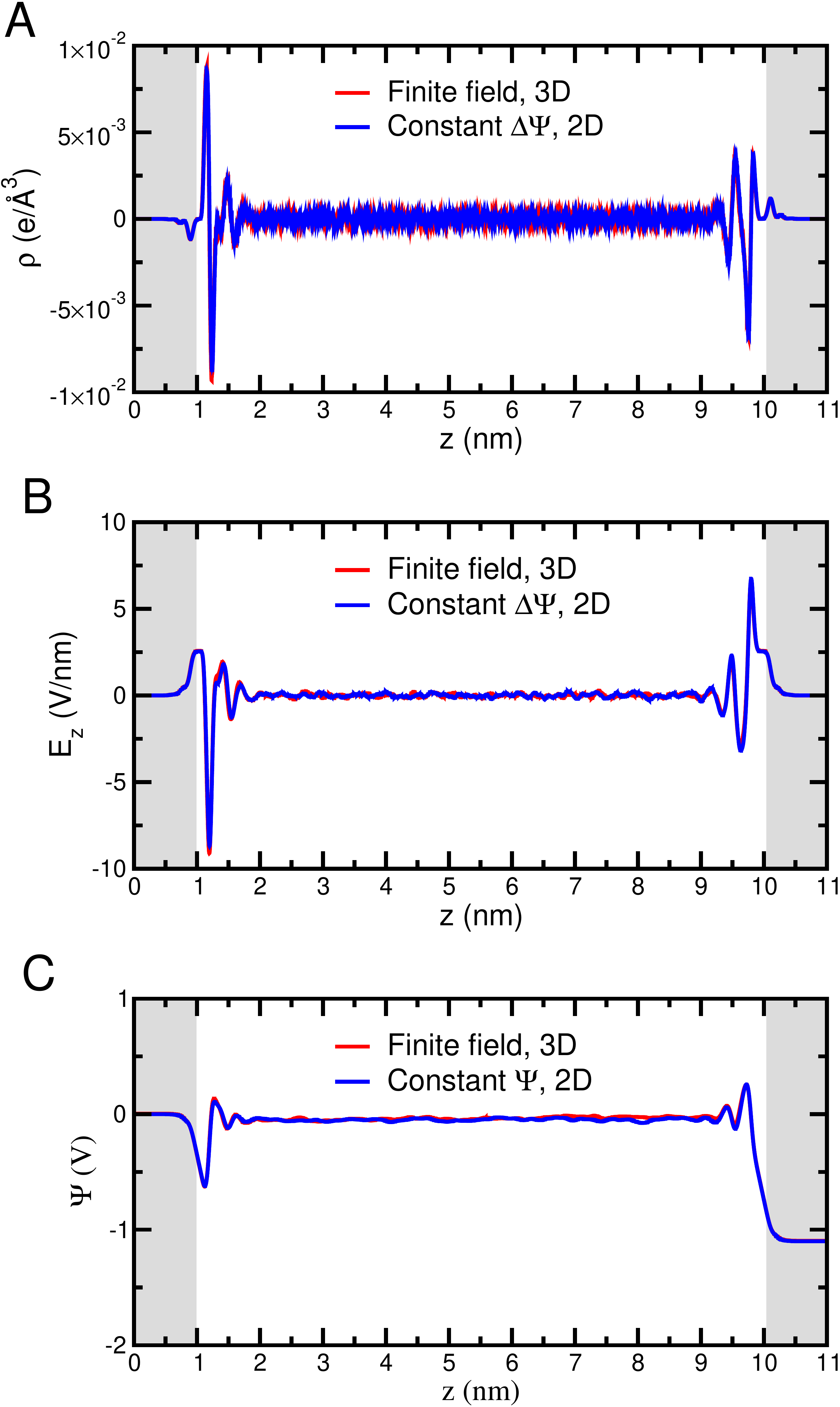}
\end{center}
\caption{Charge density (A), electric field (B) and Poisson potential (C) profile for an applied field of 0.1V/nm (corresponding to a potential drop across the cell of 1.1~V. The gray-shaded area corresponds to the electrodes (the CCS was translated in order to match with the ECS setup).\label{field_electrolyte}}
\end{figure}

As seen in the main text, fixed field and constant potential agree very well on the charge distribution (Fig \ref{field_electrolyte} A). From here one obtains the electric field profile in the z direction, averaged on the x and y directions, using the Maxwell-Gauss equation
\begin{equation}
\dfrac{\mathrm{d}E_{z}}{\mathrm{d}z}=4\pi\rho(z),
\end{equation}

with the condition $\langle E_{z}\rangle=E$, the applied field. Fig \ref{field_electrolyte} B shows that both methods agree once again. In particular, they both display a null electric field in the bulk electrode and electrolyte as expected for conductors. This is highlighted when we calculate from the electric field profile the Poisson potential profile displayed on fig \ref{field_electrolyte} C, since the potential is flat in the electrode for both methods. We also observe the same potential drop for both methods between the electrode and the electrolyte at each of the interfaces. Thus the capacitance of each half-electrode in the 3D case is the same as the 2D equivalent electrode. The agreement between applied field and constant potential is thus extended to the Poisson potential.

\subsection*{Field profiles for the single elctrode}

\begin{figure}[!h] \centering
\begin{center}
\includegraphics[width=0.4\textwidth]{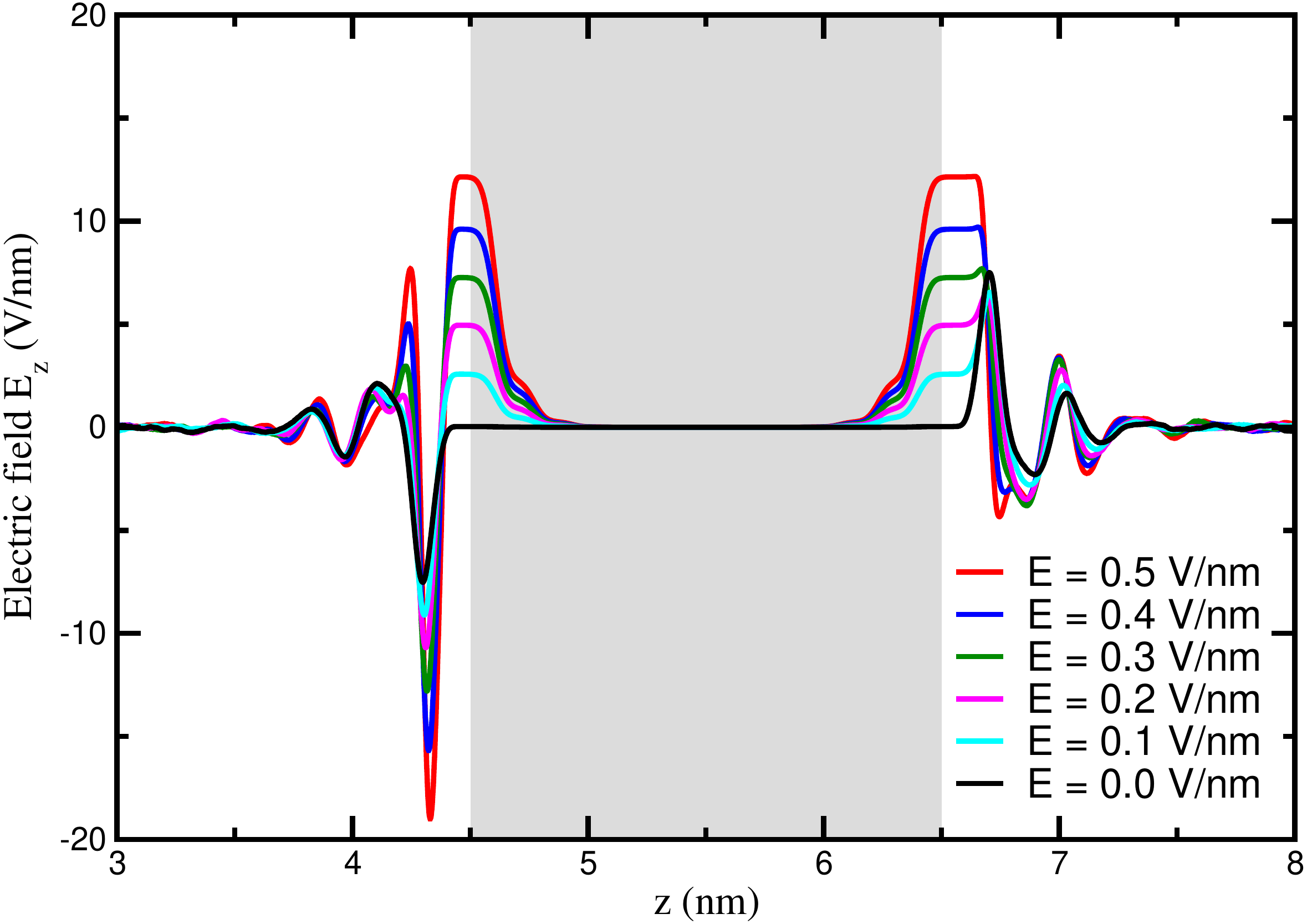}
\end{center}
\caption{Electric field profile for the single electrode under an applied electric field (corresponding to potential drop across the cell of 0 to 5.5~V). The gray-shaded area corresponds to the electrode.\label{electric_field}}
\end{figure}

One feature we need to check is if the single electrode under an applied field does behave like a perfect metal. This implies that the electric field in the bulk of the electrode is really null for every applied electric field. The electric field profile is displayed for the single electrode for multiple values of the applied electric field on fig \ref{electric_field}.

\begin{figure}[!h] \centering
\begin{center}
\includegraphics[width=0.4\textwidth]{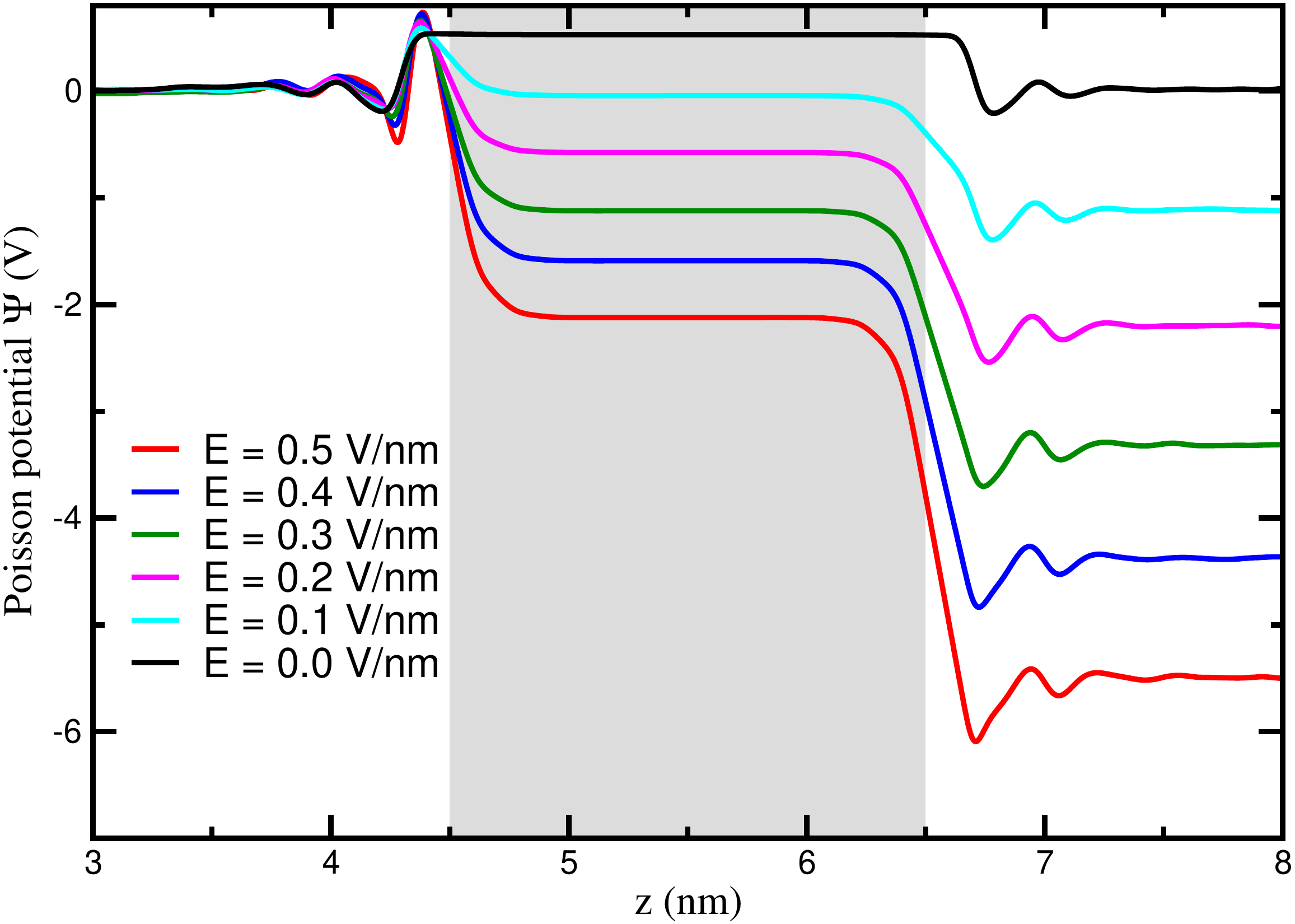}
\end{center}
\caption{Poisson potential profile for the single electrode under an applied electric field (corresponding to potential drop across the cell of 0 to 5.5~V). The gray-shaded area corresponds to the electrode.\label{Poisson_potential}}
\end{figure}

Whatever the applied field, the field in the bulk electrode is equal to zero after the second atomic layer. A closer look on this field may be obtained by the calculation of the Poisson potential. This profile is displayed for the single electrode for multiple values of the applied electric field on fig \ref{Poisson_potential}.

In the electrode, the Poisson potential profile is flat, corresponding indeed to a zero electric field in the electrode, which does behave like a perfect metal. The plateau observed allows to assign a value of the Poisson potential to the single electrode. We also notice that because of the charge density fluctuations in the electrolyte, the potential difference $\Delta\Psi$ between the two bulk part of the electrolyte on each side of the electrode is not strictly equal to $-EL_{z}$. This slight difference is otherwise not significant and is lower than the fluctuations of $\Delta\Psi$.

\subsection*{Fluctuation-dissipation relation}
\begin{figure}[!h] \centering
\includegraphics[width=0.4\textwidth]{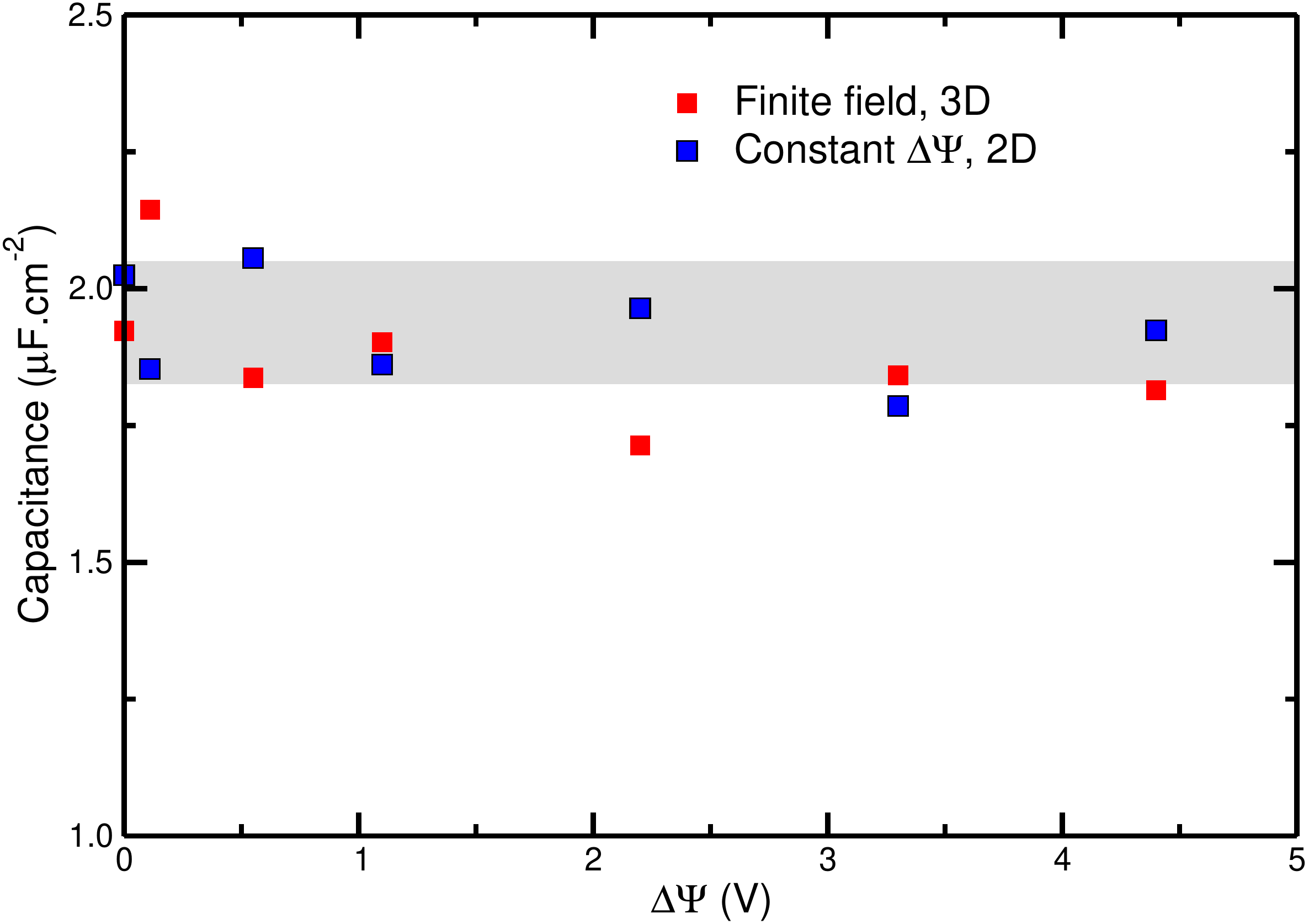}
\caption{Capacitance obtained using the fluctuation-dissipation theorem for the finite electric field (red) and constant potential (blue). The grey-shaded area corresponds to the confidence interval obtained through the $Q(\Delta\Psi)$ relation\label{fluctuation_dissipation}}
\end{figure}

The capacitance of the system sudied in the main text may be obtained from the $Q(\Delta\Psi)$ relation displayed in the main text, and gives a value of $C=1.94 \pm 0.11 ~\mu  \mathrm{F.cm}^{-2}$. It may also be calculated using the fluctuation-dissipation theorem:
\begin{equation}
C=\dfrac{1}{S}\dfrac{\langle Q^{2}\rangle - \langle Q\rangle^{2}}{k_{B}T},
\end{equation}

where $S$ is the cross-section and $T$ the temperature. The obtained values and a comparison with the fit of the $Q(\Delta\Psi)$ relation are displayed on figure \ref{fluctuation_dissipation}, showing that the two approaches are equivalent. This is not surprising since we observed in the main text that the charge distributions for constant electric field and constant applied potential are very similar in terms of both average values and standard deviation. This confirms that the capacitance does not depend on the applied voltage in the range [0:5V], that we could infer from $Q(\Delta\Psi)$ relation.

\section*{Acknowledgments}

 This project has received funding from the European Research Council (ERC) under the European Union's Horizon 2020 research and innovation programme (grant agreement No. 771294).

\bibliographystyle{apsrev}
\bibliography{references}
\end{document}